# Thermal Boundary Resistance Measurement and Analysis Across $SiC/SiO_2$ Interface

Shichen Deng [1,2,#], Chengdi Xiao [1,3,#], Jiale Yuan[2], Dengke Ma[2,4], Junhui Li[1], Nuo Yang[2,*], Hu He[1,*]

*1 State Key Laboratory of High Performance Complex Manufacturing, College of Mechanical and Electrical Engineering, Central South University, Changsha 410083, P. R. China*

*2 State Key Laboratory of Coal Combustion and School of Energy and Power Engineering, Huazhong University of Science and Technology, Wuhan 430074, P. R. China*

*3 School of Mechatronics Engineering, Nanchang University, Nanchang 330031, P. R. China*

*4 NNU-SULI Thermal Energy Research Center (NSTER) & Center for Quantum Transport and Thermal Energy Science (CQTES), School of Physics and Technology, Nanjing Normal University, Nanjing 210023, P. R. China*

# The authors contributed equally to this work.

* Corresponding to: nuo@hust.edu.cn and hehu.mech@csu.edu.cn

## Abstract

Silicon Carbide (SiC) is a typical material for third-generation semiconductor. The thermal boundary resistance (TBR) of 4H-$SiC/SiO_2$ interface, was investigated by both experimental measurements and theoretical calculations. The structure of 4H-$SiC/SiO_2$ was characterized by using transmission electron microscopy and X-ray diffraction. The TBR is measured as $8.11\times10^{-8}$ $m^2K/W$ at 298 K by 3ω method. Furthermore, the diffuse mismatch model was employed to predict the TBR of different interfaces which is in good agreement with measurements. Heat transport behavior based on phonon scattering perspective was also discussed to understand the variations of TBR across different interfaces. Besides, the intrinsic thermal conductivity of $SiO_2$ thin films (200~1,500 nm in thickness) on 4H-SiC substrates was measured by 3ω procedure, as 1.42 W/m·K at 298 K. It is believed the presented results could provide useful insights on the thermal management and heat dissipation for SiC devices.

In recent years, microelectronics industry has made great achievements in promoting the electronic performance of the semiconductors. However, due to the increasing power density and decreasing characteristic size of electronic devices, heat dispassion has become one of the most important challenges to the performance and reliability of the devices. When the device size is reduced to length scales on the order of the mean free path of the energy carrier, thermal transport is mostly determined by the thermal conductance at the interfaces between adjacent materials, rather than the intrinsic thermal properties of the materials [1-3]. Thus, it is an urgent issue to investigate the thermal boundary resistance (TBR), i.e., thermal interface resistance, in electronic devices.

Silicon carbide (SiC) is widely recognized as one of the most promising semiconductor materials for new generations of power devices due to the unique advantages in terms of power conversion efficiency, thermal conductivity and robust mechanical properties [4-6]. Moreover, SiC is the only known wide-bandgap semiconductor with a native $SiO_2$ perpetrated in the same way as silicon, which makes it suitable for metal-oxide-semiconductor devices [7]. Although SiC devices can operate at much more aggressive thermal conditions, such as higher junction temperature, their performance can also be limited by poor heat dissipation owing to the existence of TBR [8]. Particularly, the interface between silicon carbide and silicon dioxide ($SiC/SiO_2$) is the most common interface for current SiC-based microelectronics devices. Therefore, the TBR of $SiC/SiO_2$ interface is critical in thermal analysis and thermal management in SiC devices.

Recently, several researchers have studied TBR of selected heterogeneous interfaces via numerical methods, theoretical models as well as experimental methods. For numerical method, molecular dynamics (MD) simulation is widely used to predict TBR. By using MD simulation, Chen et al. studied TBR across $Si/SiO_2$ [9], Yang et al. studied TBR across Al/Si interface [10], and Hahn et al. studied TBR across Si/Ge interface [7]. The MD results are of enough accuracy and can often be consistent with experimental data [11, 12]. However, in MD simulation, constructing a detailed interface model between two different materials that have high density of interface states [13] is a challenging and sophisticated work. For theoretical models, the acoustic mismatch model (AMM) and the diffuse mismatch model (DMM) [14] have been widely used to predict and evaluate the TBR. Both models make simplified assumptions of the phonon scattering processes across interface [15], and some research works have been carried out to improve these two models [16-18]. Compared with the complexity and time consumption of MD simulations, AMM or DMM can avoid the complex interface modeling yet achieve reasonable prediction on TBR. Thus, the AMM and DMM are easier to employ and then predict the TBR qualitatively and efficiently. For experimental methods, time-domain thermoreflectance method (TDTR) [19] and 3ω method [20] are commonly used to study the interfacial thermal transport at micro-nano scale. Lyeo et al. measured the TBR across Pb/Si interface using TDTR [21], Wang et al. measured the TBR across $SiO_2$/GaN interface using 3ω method [22], and Kuzmik et al. measured the TBR across GaN/Si interface with Micro-Raman method [23].

However, the thermal behavior of SiC/$SiO_2$ interface is lack of investigation. Numerous studies have shown that there is a high density of interfacial traps between SiC and its native oxide [24-26], which leads a lower electron mobility that affects the performance of SiC devices. It is still unknown how much the roughness would block phonon transport across SiC/$SiO_2$ interface. The related results indicated that the interfacial thermal conduction strongly depends on defect density at interface. English et al. found that compositional disorder at an interface could strongly reduce thermal interface conductance at high temperature (50% of melting temperature) [27]. Liang et al. investigated the effect of interfacial parameters on thermal accommodation coefficient (TAC) at solid-gas interface, which revealed that the TAC on smooth and perfect interface is significantly lower than that on a disordered interface [28]. This can be attributed to the roughness and defects on the disordered interface, which is more conducive to diffuse scattering of gas molecules. Therefore, it is indispensable to determine the TBR across SiC/$SiO_2$ interface, which is critical for the efficient design and thermal management of SiC devices.

In this work, the thermal boundary resistance of 4H-SiC/$SiO_2$ interface is investigated by both experimental 3ω method and theoretical diffuse mismatch model. In order to verify the effectiveness of the customized 3ω experimental platform, the TBR of the Si/$SiO_2$ interface has been investigated. Then, the thermal conductivities of $SiO_2$ films with various thicknesses (200 nm ~ 1500 nm) on 4H-SiC substrates have been measured. Experimental characterization and theoretical analysis have been conducted to determine the results of TBR across 4H-SiC/$SiO_2$ interface. Therefore, high credibility and better reproducibility of the TBR data can be obtained.

In order to obtain the TBR between 4H-SiC/$SiO_2$ interface by the 3ω method, five different thicknesses of $SiO_2$ film were deposited by Plasma Enhanced Chemical Vapor Deposition (PECVD) on the Si and 4H-SiC substrate, respectively. Additionally, the feasibility of our customized 3ω measurement platform was demonstrated for the TBR of Si/$SiO_2$ interface and the thermal conductivity of $SiO_2$ thin film with respect to previous reported results in the literature. Details on sample preparation, measurement process as well as characterization methods are addressed in supporting information (SI).

In the following, the measurement of 4H-SiC/$SiO_2$ interface was conducted by the customized 3ω measurement platform. Table 1 lists the measured values of the apparent thermal conductivities and apparent thermal resistances of the $SiO_2$ thin films as a function of film thickness. The apparent thermal conductivity values of the $SiO_2$ thin film samples with thicknesses varied from 200 to 1500 nm are between 0.96 and 1.32 W/m·K at room temperature, 298 K. The apparent thermal resistances of the $SiO_2$ thin films on the 4H-SiC substrate are plotted in Fig. 1.

**Table 1. Experimental results of the apparent thermal conductivities and apparent thermal resistance of the $SiO_2$ thin films with different thicknesses.**

| Thickness (nm) | $\kappa_f$ (W/m·K) | $R_f$ ($\times10^{-8}$ m$^2$K/W) |
|---|---|---|
| 200 | 0.96±0.14 | 21.54±3.18 |
| 400 | 1.02±0.12 | 39.94±5.49 |
| 500 | 1.16±0.04 | 42.97±1.35 |
| 700 | 1.22±0.02 | 57.41±1.05 |
| 1000 | 1.28±0.03 | 78.06±2.08 |
| 1500 | 1.32±0.03 | 113.43±1.37 |

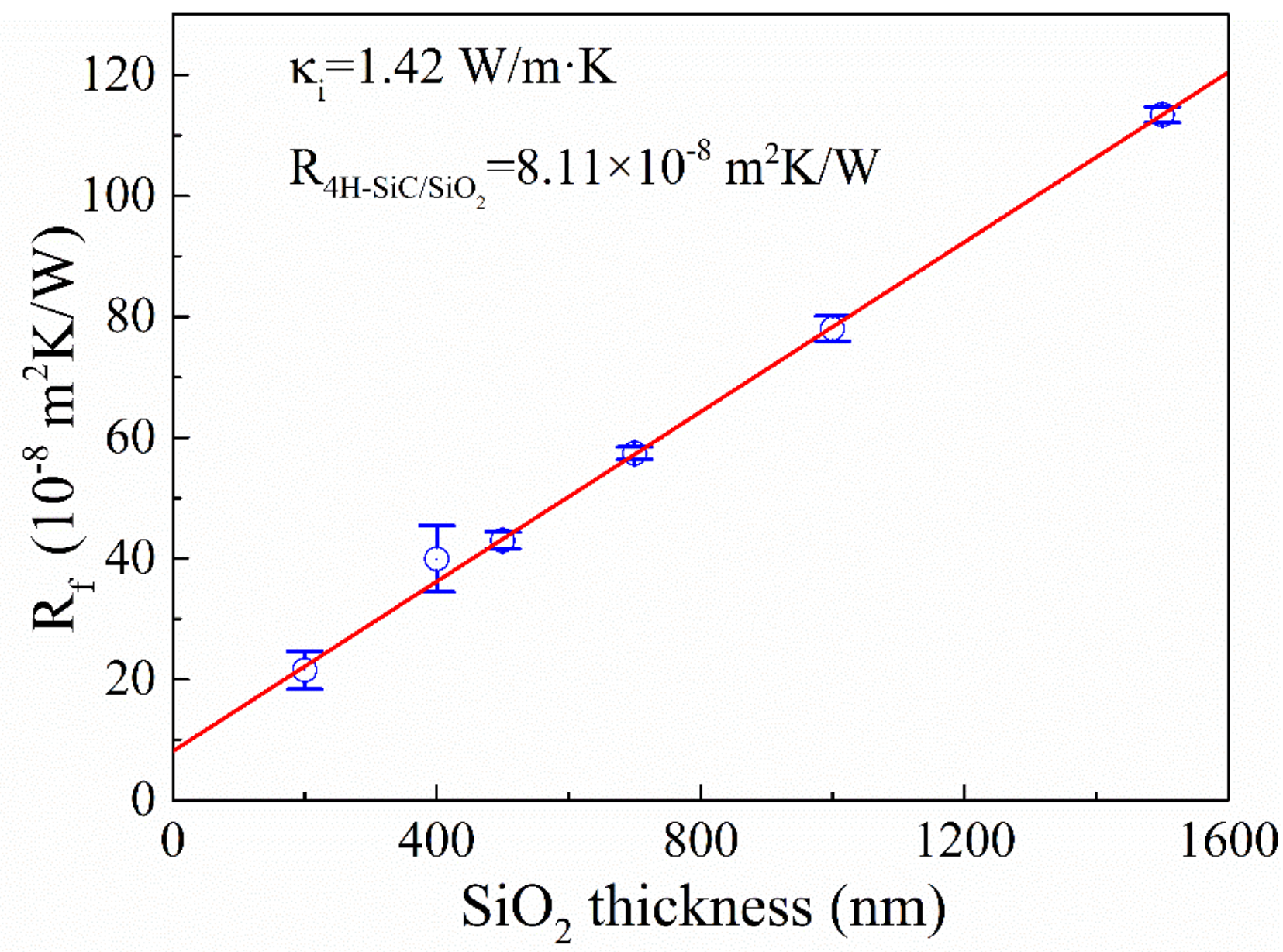

**FIG. 1.** Apparent thermal resistance ($R_f$) of $SiO_2$ thin films on 4H-SiC substrate. The red line is the linear fitting of the measured data.

The intrinsic thermal conductivity of $SiO_2$ thin film on 4H-SiC substrate was obtained as 1.42 W/m·K, which is slightly greater than that of $SiO_2$ film on Si. This can be attributed to the preparation of two films, which are deposited separately in different PECVD equipment. The TBR of 4H-SiC/$SiO_2$ interface was calculated as $8.11\times10^{-8}$ $m^2K/W$ at 298 K, which is about 4 times higher than the value of Si/$SiO_2$. A larger TBR of 4H-SiC/$SiO_2$ could be due to the fact that there is high density of imperfections at 4H-SiC/$SiO_2$ interfaces, which restrains phonon heat transport across interface. Later, TEM testing and theoretical analyses will be conducted to explain the discrepancy between these two interfaces. The TBR of 4H-SiC/$SiO_2$ interface measured in our work is the same order of magnitude as the values of similar heterogeneous interface system presented in literature. The previously reported TBR between highly dissimilar materials at room temperature measured using TDTR, ranges from $3.33\times10^{-8}$ to $1.25\times10^{-7}$ $m^2K/W$ [21]. Kuzmik et al. measured TBR of the GaN/SiC interface by Micro-Raman method to be about $1.2\times10^{-7}$ $m^2K/W$ [23].

In order to furtherly investigate the TBR of 4H-SiC/$SiO_2$ interface, Fig. 2(a) shows the interfacial structure at the 4H-SiC/$SiO_2$ interface. The interface structure is an important factor to heat transport across heterogeneous interface, which affects the phonon transport and scattering at interfaces. Potential damage at the near-surface region of 4H-SiC substrate can be observed, and the disordered 4H-SiC layer has a thickness of 3~5 nm. Previous studies have demonstrated that structural disorder would lead to an increase of the TBR due to the increased number of phonons scattering at the disordered region [29]. In addition, it is observed at the 4H-SiC/$SiO_2$ interface close to the side of amorphous $SiO_2$ that there are multi-vacancy defects induced by threefold coordinated O and C interstitial atoms and lattice mismatch of 4H-SiC and $SiO_2$ materials [26]. On one hand, these vacancy defects would reduce the effective contact area compared with smooth contacting surfaces, and the poor contact between two materials would predominantly increase the TBR. On the other hand, these vacancy defects would increase the phonon scattering with defects and reduce phonon transmission at the interface, which eventually would increase the TBR. Generally, the interface defect density at Si/$SiO_2$ interface remains one to two orders of magnitude lower than that typically found at 4H-SiC/SiO2 interface [30], which contributes to a lower value of TBR of Si/$SiO_2$.

On the other side, the thermal conductance could increase due to rough interface in certain cases. For instance, MD simulations have shown that TBR of graphene/copper decreases with the increase of roughness [31], which arises from the high local pressure between graphene and copper substrate. Besides, it is found that interfacial roughness could decrease TBR due to the increased interface areas and the "phonon bridging" effect [32], where two materials are assumed to contact with each other extremely well in MD simulations. Thus, TBR can be reduced by the rough interfaces due to increase the contact strength or effective contact area, which is not valid in samples of this work.

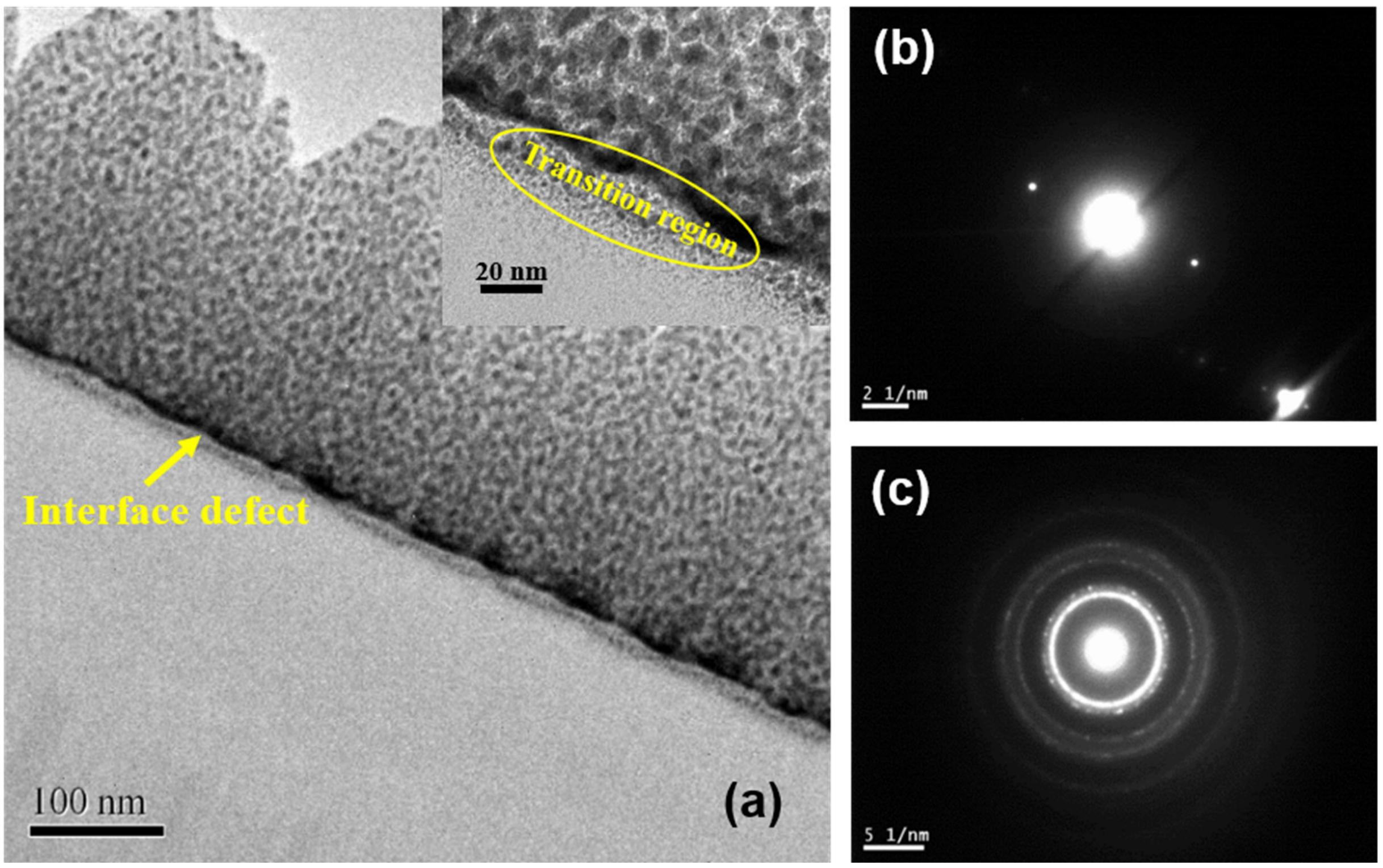

**FIG. 2.** The cross-sectional TEM image of the 4H-SiC/$SiO_2$ interface and SAED pattern on both sides of 4H-SiC/$SiO_2$ interface. (a) Interface TEM image with 100nm in scale. The inset represents the magnified region with 20nm in scale; (b) 4H-SiC SAED pattern; (c) $SiO_2$ SAED pattern.

Meanwhile, selected area electron diffraction (SAED) was employed to determine the state of the materials on both sides of the 4H-SiC/$SiO_2$ interface. Fig.2 (b)-(c) show that 4H-SiC substrate is in a typical crystalline state, while $SiO_2$ film is polycrystalline. Generally, the thermal conductivity of the crystalline material is much higher than that with amorphous state, and the thermal conductivity of the bulk crystal $SiO_2$ is about 10.4~11.3 W/m·K. Therefore, the higher thermal conductivity of the $SiO_2$ film on the 4H-SiC substrate can be interpreted in terms of polycrystalline state.

Afterwards, theoretical analyses were conducted to both interfaces. Since the roughness of experimentally manufactured interface is usually large, the commonly used diffuse mismatch model (DMM) [14, 33] is chosen to analyze the thermal boundary resistance of Si/$SiO_2$ and 4H-SiC/$SiO_2$ interfaces. Considering an interface between two materials, namely, A and B, a phonon with frequency ω and mode j that is incident on the interface from A can either scatter back into A or transmit into B. The DMM assumes that scattering at the interface is completely diffusive in nature. In other words, an incident phonon at the interface loses memory of its original state and scatters into a phonon state with the same energy in either medium. The probability of a phonon diffusely scattering across the interface is related to the density of states and the phonon group velocity on each side of the interface. Under the assumption of diffusive scattering, as well as invoking the principle of detailed balance [14], the transmission coefficient from material A to material B, $\alpha_{A\rightarrow B}$, can be calculated as:

$$\alpha_{A\rightarrow B}=\frac{\sum_j D_{B,j}v_{B,j}}{\sum_j D_{A,j}v_{A,j}+\sum_j D_{B,j}v_{B,j}} \tag{1}$$

where $D$ is the phonon density of states, $v$ is the phonon group velocity for the phonon mode of interest at frequency ω, the subscript "$j$" refers to the phonon polarization.

According to the Landauer formula, the interface thermal conductance (G) can be predicted as

$$G=\frac{1}{4}\sum\int_0^{\omega_A^v} D_A(\omega)\frac{\partial n(\omega,T)}{\partial T}\hbar\omega v_A\alpha_{A\rightarrow B}(\omega)d\omega \tag{2}$$

where $\omega^{\nu}$ is the cut-off frequency, $n(\omega,T)$ is the Bose-Einstein distribution function.

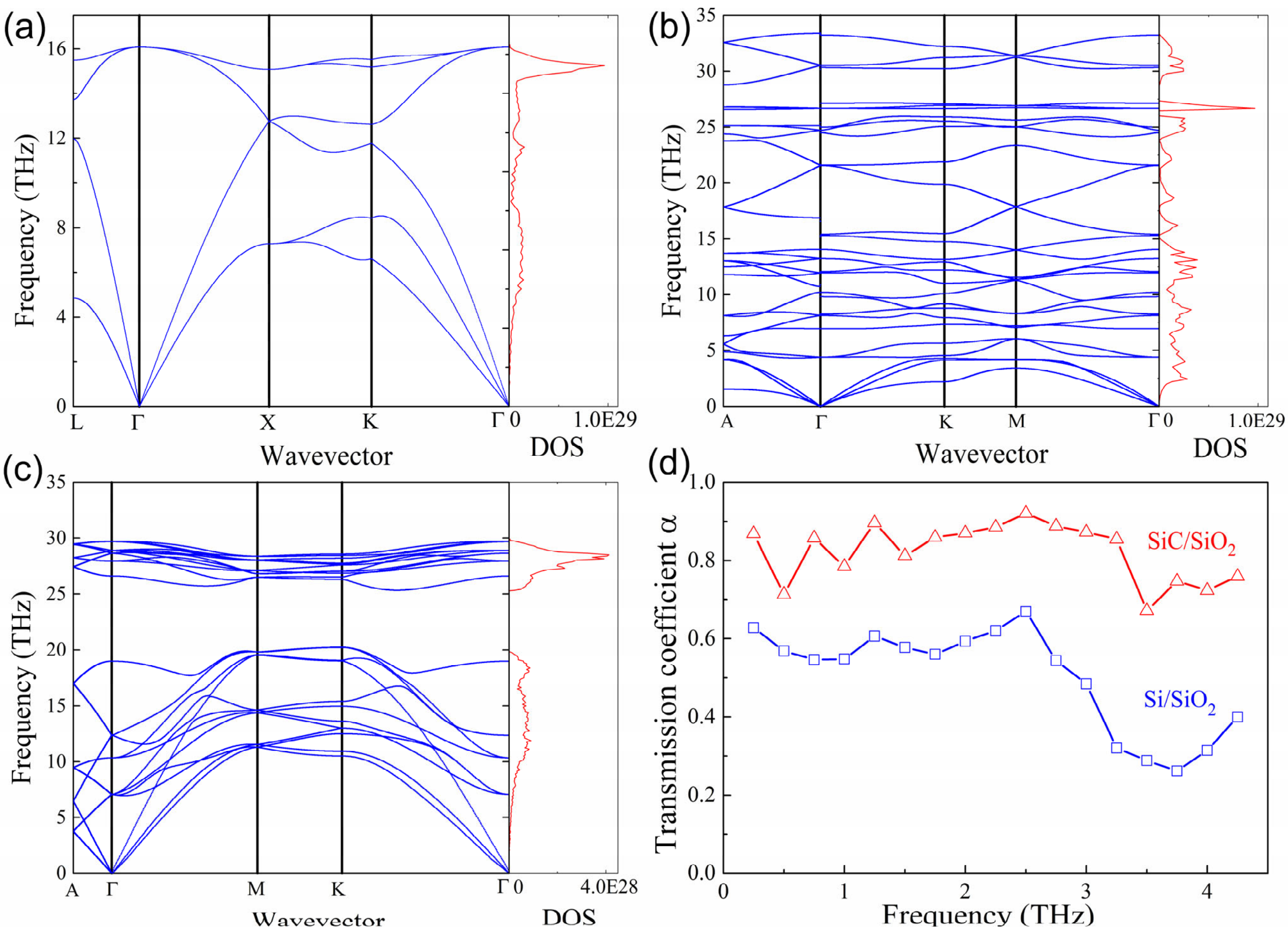

**FIG. 3.** Phonon dispersions and density of states (DOS) of (a) Si, (b) $SiO_2$ and (c) 4H-SiC, respectively. (d) The transmission coefficient (α) of $Si/SiO_2$ and 4H-$SiC/SiO_2$ interfaces.

Fig. 3(a)-(c) show the phonon dispersions and density of states for Si, 4H-SiC and $SiO_2$ crystals calculated with GULP [34] code, respectively. The phonon group velocities can be derived from phonon dispersions. Tersoff potentials are applied for Si [35] and 4H-SiC [36], while core-shell spring potential is used for $SiO_2$ [37]. The calculation results are in good agreement with other results from first-principle theory [38-40]. Fig. 3(d) shows the phonon transmission coefficients of $Si/SiO_2$ and 4H-$SiC/SiO_2$ interfaces obtained from DMM. The transmission coefficient α of phonons in 4H-SiC to $SiO_2$ is larger than that of Si. This is because the discrepancy of phonon DOS between 4H-SiC and $SiO_2$ is higher than that between Si and $SiO_2$. As illustrated in Eq. (1), the higher the discrepancy of phonon DOS, the larger the transmission coefficient would be.

**Table 2. The thermal boundary resistances. (Unit: $m^2K/W$)**

| Samples | Theoretical Prediction | Experimental Measurement |
|---|---|---|
| $Si/SiO_2$ | $6.88\times10^{-9}$ (298 K) | $2.13\times10^{-8}$ (298 K) |
| 4H-SiC/$SiO_2$ | $2.72\times10^{-8}$ (298 K) | $8.11\times10^{-8}$ (298 K) |
| Si/Al (DMM) | $3.70\times10^{-9}$ (300 K) [41] | / |
| $Si/SiO_2$ (MD) | $4.27\times10^{-9}$ (300 K) [9] | / |
| Si/Graphene (MD) | $4.3\times10^{-8}$ (300 K) [42] | / |
| Si/GaN (Micro-Raman) | / | $7.5\times10^{-8}$ (420 K) [23] |
| SiC/GaN (Micro-Raman) | / | $1.2\times10^{-7}$ (420 K) [23] |

Acoustic phonons were considered to calculate thermal boundary resistances [41]. The predicted TBR along with experiment results were presented in Table 2. The DMM value for $Si/SiO_2$ interface is $6.88\times10^{-9}$ $m^2K/W$, while for 4H-SiC/$SiO_2$ interface the value is $2.72\times10^{-8}$ $m^2K/W$ at 298 K. Our calculation results are close to other theoretical or simulation predictions for other similar interfaces listed in Table 2. Reddy et al. calculated the TBR values of Al/Si interface as $3.70\times10^{-9}$ $m^2K/W$ at 300 K with DMM [41]. Chen et al. showed that the TBR value of $Si/SiO_2$ is about $4.27\times10^{-9}$ $m^2K/W$ in weak interfacial coupling cases with MD [9]. Shen et al. showed that the TBR of Si/Graphene interface is about $4.3\times10^{-8}$ $m^2K/W$ with MD [42].

Additionally, compared the predicted TBR values by DMM with the experimental results, it was found that the experimental results are two to three folds of the predicted values. This discrepancy can be interpreted by considering the limitations and assumptions made in DMM. There are three main mechanisms that DMM could undervalue TBR. Firstly, when calculating phonon dispersion relation by using lattice dynamics, the material structure is assumed to have a perfect crystallinity. Whereas, the samples of $SiO_2$ film in this work are amorphous. There are obvious differences in PDOS between amorphous structure and crystallized structure. For instance, due to much more scatterings, the peaks in PDOS of amorphous are widened and are not as sharp as that of crystal structure [43]. The overlapping region of PDOS is upvalued when using the crystal structure, which also upvalues the phonon transmission probability across the interface, thus undervalue TBR. Secondly, DMM assumes perfectly contacted rough scattering interfaces with no imperfections. That is, it cannot assess the effects of defects and disorder, which are generally existed at interface. While, as shown in Fig. 2, defects and disorder at the near-surface region of 4H-SiC substrate are observed. These defects and disorders make the contact area smaller than that of smooth contacting surfaces [44], and also induce more scatterings of phonons [29]. It contributes to a larger TBR of experimental results, and similar results are observed in Au-Si [45] and Al-Si [29] interface. Thirdly, DMM inherently assumes diffusive thermal transport across interfaces, which might not be always true. It is shown that DMM could underestimate TBR when there are nondiffusive phonon scatterings at interfaces by using both theoretical calculations and molecular dynamics [46]. Therefore, the assumption of totally diffusive scattering in DMM is potentially inevitable to contribute the discrepancy between computational value and experimental value. On the other side, another limitation of DMM may overestimate TBR. That is, DMM does not take account for the inelastic scattering of phonons at interfaces, while inelastic scatterings would provide more channels for phonon transport and reduce TBR.

Moreover, both the predicted and experimental results demonstrate that 4H-SiC/$SiO_2$ interface has a higher thermal resistance than Si/$SiO_2$ interface. On one hand, the different phonon DOS between 4H-SiC and Si attributes to the difference of TBR between heterogeneous interfaces. As in Eq. (1), the interfacial thermal conductance is influenced by phonon DOS, phonon group velocity and the transmission coefficient. While the phonon density of states and group velocity are the intrinsic properties of the material, the transmission coefficient is determined by both materials across the interface (Eq. (S6) in the supporting information (SI)). As shown in Fig. 3(d), the highest frequency of the acoustic phonons in $SiO_2$ is 4.2 THz, which means phonons in 4H-SiC with a frequency higher than 4.2 THz are restricted to transport through the interface, thus, only the transmission coefficients of phonons with frequencies below 4.2 THz are calculated. Our calculation results show that 4H-SiC/$SiO_2$ interface has a slightly higher phonon transmission efficiency than Si/$SiO_2$ interface in general, and the group velocity of 4H-SiC is about 1.5 times of that of Si. However, the phonon DOS of 4H-SiC is about an order of magnitude lower than that of Si when the frequency is below 4.2 THz. The lower phonon DOS of 4H-SiC in low frequency can be attributed to the lighter atomic mass of C compared to Si, which causes the 4H-SiC lattice has more vibration modes on higher frequency [47]. Therefore, lower phonon DOS of 4H-SiC at frequency below 4.2 THz implicates that heat carriers that can pass through the 4H-SiC/$SiO_2$ interface is less than the Si/$SiO_2$ interface, causing higher TBR in 4H-SiC/$SiO_2$ interface. On the other hand, when a film is grown on the substrate with higher thermal conductivity, the influence of TBR on the heat transport might be more significant. Kuzmik et al. [23] measured the TBR of Si/GaN and SiC/GaN interfaces using Micro-Raman spectroscopy technique, which are listed in Table 2. The TBR of Si/GaN interface is $7.5\times10^{-8}$ $m^2K/W$ while the TBR of SiC/GaN interface is $1.2\times10^{-7}$ $m^2K/W$, which also supports the address that SiC substrate with higher thermal conductivity introduces larger TBR of SiC/GaN interface comparing to that of Si/GaN interface.

In summary, the thermal boundary resistance of both 4H-SiC/$SiO_2$ and Si/$SiO_2$ interfaces were measured and compared by 3ω method. It was found the values of TBR between $SiO_2$ thin film and 4H-SiC substrate is $8.11\times10^{-8}$ $m^2K/W$, which is about 4 times higher than that of Si/$SiO_2$ interface. The TBR values of these two interfaces are within the same range with the previously reported TBR between other highly dissimilar materials at room temperature measured using TDTR which ranges from $3.33\times10^{-8}$ to $1.25\times10^{-7}$ $m^2K/W$ [21], as well as the results of GaN/SiC interface by Micro-Raman method about $1.2\times10^{-7}$ $m^2K/W$ [23].

Additionally, the diffuse mismatch model was deployed to calculate and analyze the TBR of 4H-SiC/$SiO_2$ and Si/$SiO_2$ interfaces, the results is $2.72\times10^{-8}$ and $6.88\times10^{-9}$ $m^2K/W$, respectively, which is well match to our experimental results. Based on the microstructure characterization and theoretical analysis for different interfaces, the discrepancy between two interfaces were attributed to two potential explanations: 1) There is higher density of imperfections at 4H-SiC/$SiO_2$ interfaces which affect phonon heat transport across interface significantly; 2) The DOS of low frequency acoustic phonons in 4H-SiC is lower than that of Si, leading less heat carriers in 4H-SiC transport through the interface. The presented work on TBR of 4H-SiC/$SiO_2$ interface is believed to pave a way for heat dissipation, thermal analysis and thermal management in high power density devices.

## Supporting Information

Sample preparation, measurement process, characterization methods, heat transport across Si/$SiO_2$ interface, apparent thermal conductivities of the $SiO_2$ thin films on the Si substrate at 298K, apparent thermal resistance of $SiO_2$ thin films on Si substrate, energy dispersive spectrometer (EDS) characterization for 4H-SiC/$SiO_2$ interface, and details of numerical calculations are included.

## Acknowledgements

H.H. was sponsored by National Program on Key Basic Research Project (2015CB057206), National Natural Science Foundation of China (51605497) and the State Key Laboratory of High Performance Complex Manufacturing (ZZYJKT2019-05). N.Y. was sponsored by National Natural Science Foundation of China (No. 51576076 and No. 51711540031), Hubei Provincial Natural Science Foundation of China (2017CFA046) and Fundamental Research Funds for the Central Universities (2019kfyRCPY045). The authors thank the National Supercomputing Center in Tianjin (NSCC-TJ) and China Scientific Computing Grid (ScGrid) for providing assistance in computations.